\documentclass[twocolumn,english,aps]{revtex4-1}
\usepackage[T1]{fontenc}
\usepackage[latin9]{inputenc}
\usepackage{array}
\usepackage{longtable}
\usepackage{multirow}
\usepackage{graphicx}

\makeatletter

\providecommand{\tabularnewline}{\\}

\makeatother

\usepackage{babel}
\begin{document}

\title{Developing a project-based computational physics course grounded
in expert practice}

\author{Christopher Burke}

\affiliation{Department of Physics and Astronomy, Tufts University, 574 Boston
Avenue, Medford, MA 02155}

\author{Timothy J. Atherton}

\email{timothy.atherton@tufts.edu}

\selectlanguage{english}%

\affiliation{Department of Physics and Astronomy, Tufts University, 574 Boston
Avenue, Medford, MA 02155}
\begin{abstract}
We describe a project-based computational physics course developed
using a backwards course design approach. From an initial competency-based
model of problem solving in computational physics, we interviewed
faculty who use these tools in their own research to determine indicators
of expert practice. From these, a rubric was formulated that enabled
us to design a course intended to allow students to learn these skills.
We also report an initial implementation of the course and, by having
the interviewees regrade student work, show that students acquired
many of the expert practices identified. 
\end{abstract}
\maketitle

\section{Introduction}

Computers have been used to solve physics problems since they were
first created\cite{haigh2016eniac}, and with their ever increasing
ubiquity, computing has become a fundamental scientific practice in
Physics. Some brief examples\cite{mccurdy2002computation}: in High
Energy Physics, computers automatically select detected events for
recording, store them and allow processing of the results on large-scale
compute Grids; moreover, theoretical predictions are made by simulations.
In Astronomy, formation processes are simulated at multiple length
scales from star formation to galaxy evolution. In Condensed Matter
Physics, many-body quantum mechanics simulations predict the electronic
structure of crystals and molecules, bulk properties from molecular
dynamics and continuum behavior by solving PDEs. Moreover, almost
since they became available, computers have been used in Physics Education
for visualization and simulation\cite{Goldberg1967}, and their immense
potential as educational tools was recognized early on\cite{bork1979interactive,bork1979special}.
This tradition is continued today in projects such as the PhET\cite{perkins2006phet}
simulations, designed for classroom use. 

Given the central importance of computation to physics research, and
hence the need for appropriately trained students and professionals
to perform it, it is perhaps surprising that computing is often only
weakly integrated into the physics curriculum. Several approaches
have been tried to remedy this\cite{chonacky2008integrating}: At
the largest scale, whole majors on Computational Science have been
created\cite{Landau2006}; courses\cite{rebbi2008} or sequences of
courses\cite{spencer2005teaching,cook2008computation} have been created
within departments; computational projects and homework problems have
also been integrated into existing courses and laboratories, both
for majors\cite{caballero2014model} and at the introductory level\cite{serbanescu2011putting,caballero2012implementing}.
Encouragingly, a large-scale survey performed about a decade ago\cite{fuller2006numerical},
showed widespread support for these efforts. A valuable starting point
for those seeking to implement one of these approaches is a special
double volume\cite{christian2008introduction} of the American Journal
of Physics on Computation published in 2008 and including several
of the articles cited above. Of particular practical help is the extensive
Resource Letter\cite{landau2008resource}, an updated version of\cite{devries1996resource}. 

In this work, we discuss a Computational Physics class recently created
at Tufts University. Central to our vision was a desire to ground
the course in actual professional practice, following a movement to
refocus science education around real scientific practices\cite{cooper2015challenge}.
We therefore conducted interviews of scientists, mathematicians and
computer scientists to determine these practices. The results of the
these interviews were distilled into a rubric that aims to describe
indicators of expert practice in computational physics. From the rubric,
we used a backwards course design approach to create a sequence of
projects that allow students to acquire these practices in a structured
manner. We gave the course for the first time in the Spring 2015 semester,
and conducted an after-delivery assessment exercise to determine whether
student work did indeed contain evidence of expert practice. We report
in detail the design process in the hope it might be of use to others
planning project-based courses with other subject matter. 

The structure of the paper follows our design and assessment exercise:
the rubric and its development is described in section \ref{sec:Rubric};
the course design process is explained in section \ref{sec:Course-design};
brief comments on the initial implementation are given in section
\ref{sec:Implementation}; our \emph{a posteriori} analysis of the
success is presented in section \ref{sec:Analysis}. We finish with
discussion and conclusions in section \ref{sec:Conclusion}.

\section{Rubric development\label{sec:Rubric}}

The initial phase of our design process was to develop a rubric for
the course, capturing in a single document indicators of expert practice
in Computational Physics. The purpose of this was threefold: First,
and most obviously, the rubric would be a tool to help assess the
quality of student work submitted in the course. Second, by making
this document available to students during the course, we wished to
convey these excellent practices as a norm and engender a climate
of professionalism in the classroom. Finally, we anticipated that
by distilling the practices into a rubric, the resulting document
would be a powerful tool to assist in the design of the course, ensuring
that the content and assessment would be mutually aligned, and that
both would prepare students to solve computational problems in actual
physics research. 

We began by determining the typical content of Computational Physics
courses. We examined course syllabi from a range of institutions,
Boston University, University of Connecticut, East Tennessee State
University, Northeastern University and Oregon State University. From
these syllabi, and our own experiences as professionals, we proposed
an initial set of competencies that seem essential to the practice
of computational physics: \emph{Physical transcription}, the formulation
of the physical system of interest into a mathematical problem and
creation or selection of algorithm(s) to solve it; \emph{Planning},
organizing the overall program into modules and selecting appropriate
data structures; \emph{Programming}, implementation of the approach;
\emph{Visualization}, post-processing of the results and selection
of the appropriate representations to interpret them; \emph{Numerical
analysis}, assessing error and stability of the program; \emph{Physical
analysis}, relating the results of the program back to the initial
problem and determining whether the approach has provided the required
insight. 

\begin{figure}
\begin{centering}
\includegraphics{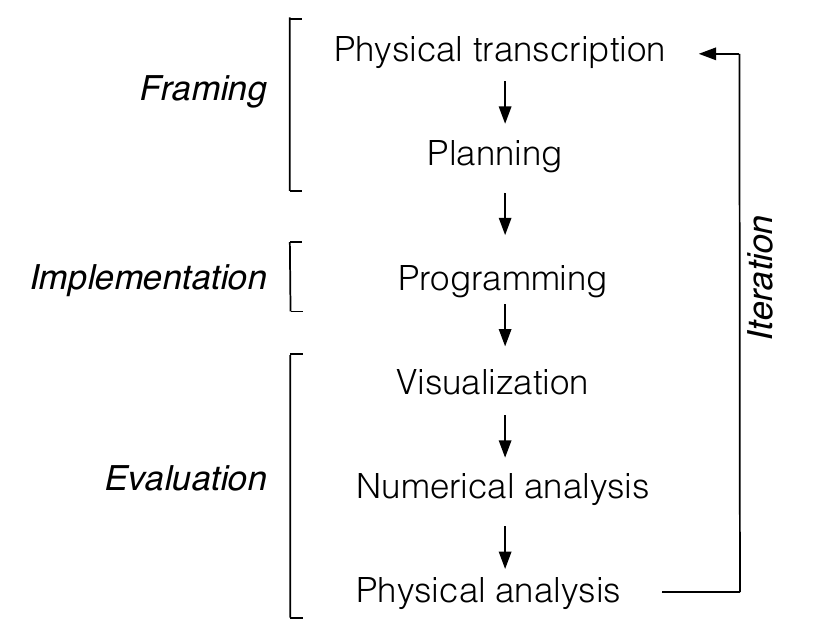}
\par\end{centering}

\caption{\label{fig:Initial-model}Initial model of expert problem solving
in Computational Physics.}
\end{figure}

Our initial model of how these competencies were used by experts to
solve problems in physics research is shown in figure \ref{fig:Initial-model}.
We anticipated a formulation phase, in which the physical system and
research questions of interest were transcribed into a math problem
and then an algorithm developed or selected to solve the problem.
Following this, the program is implemented. Having run the program,
the results are evaluated both to assess whether the results were
numerically meaningful, i.e. whether the program gave stable, repeatable
results, and physically meaningful, i.e. whether they answered the
original research questions. We envisioned that these activities would
be repeated iteratively several times, as the initial product may
not fully address the original questions, or the results may lead
to new questions. This sequence mirrors famous work on problem solving\cite{polya1957solve}
as well as physics-specific research\cite{hsu2004resource}. We particularly
note the parallel with the Integrated Problem-Solving model\cite{Litzinger2006,litzinger2010cognitive},
which divides the problem-solving process into \emph{framing, implementation
and evaluation}, and was recently adapted to assess student learning
in Computational Materials Science\cite{Vieira2015}. 

From our initial list of competencies and problem solving model, we
devised interview questions (listed in Appendix \ref{sec:Appendix-A})
to determine actual professional practices associated with these competencies.
We selected five faculty and postdocs from Physics, Computer Science
and Mathematics who actively use computation in their research and
have published refereed papers involving computation. These five were
interviewed by one of the authors (CJB) and these interviews transcribed.
The two authors then coded the transcripts separately according to
the prompt ``identifiable items we could look for in student work''.
Following this step, the coded items were compared and only common
items retained.  We then categorized the coded items according to
our initial competency model. 

In performing this exercise, we identified additional competencies
that we had not proposed in our initial model: \emph{Running}, incorporating
the ability to create scripts and file structures to execute the program
and collect the data, and \emph{Testing}, incorporating comparison
of the output of the program against known solutions as well as profiling
the program to assess performance. We renamed the competency \emph{Programming}
in the initial model to \emph{Implementation} to reflect our inclusion
of important non-programming practices associated with implementation,
e.g. creation of external documentation and use of version control.

\begin{figure}
\begin{centering}
\includegraphics{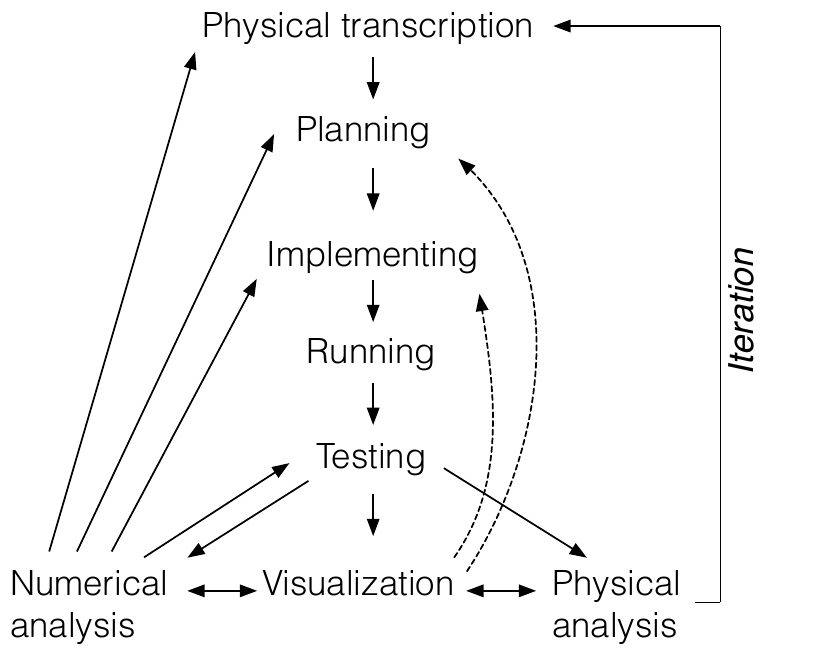}
\par\end{centering}

\caption{\label{fig:FinalModel}Evidence-based model of expert problem solving
in Computational Physics.}
\end{figure}

Furthermore, the evidence of the interviews forced us to abandon our
initial cyclic model of how Computational Physics programs are created
and instead create a more highly connected and iterative model, depicted
in fig. \ref{fig:FinalModel}. Of particular note, we found evidence
that ideas from numerical analysis inform almost every aspect of computational
physics problem solving, and that experts typically use these ideas
as a sub-step in each of the Physical Transcription, Planning, Implementing,
Testing and Visualization steps.  Testing and Numerical Analysis
proceed in a closely iterative fashion. The Numerical analysis, Visualization
and Physical Analysis steps of evaluating the program output are also
highly interconnected and mutually inform the selection of representations
to plot and how to interpret the results. Perhaps our most surprising
result is that Visualization was shown not to be limited to assessing
the program output, but also a key tool to plan and implement effective
programs. Overall, the evidence suggests that while experts do indeed
solve Computational Physics problems iteratively, the execution of
their strategy does not involve a clearly defined sequence of steps
but rather invocation of different competencies as required. While
the number of interviewees in this study is small, their nonlinear
approach to problem solving parallels that seen in other domains\cite{practice2000people}. 

A draft rubric was prepared from the coded items, organized along
the new competency model, which seeks to define expert practice through
the indicators identified from the interviews. The participants interviewed
were invited to give feedback on the rubric draft at a panel, together
with representatives from the Tufts Center for the Enhancement of
Learning and Teaching (CELT) and Tufts Technology Services Educational
Technology group who both possessed relevant expertise in computing
education. Following the recommendations of the group, we constructed
the final version of the rubric shown in table \ref{tab:Rubric}.

\begin{table*}
\begin{centering}
\begin{longtable}{|>{\raggedright}p{1in}|>{\raggedright}p{5.8in}|}
\hline 
Competency & Indicators\tabularnewline
\hline 
\hline 
Physical Transcription & Analytical methods are first employed to understand the problem to
the greatest extent possible, including identification of symmetries,
length scales and timescales. An appropriate mathematical model and
computational representation are chosen including choice of algorithm
and discretization. Multiple methods are considered and their strengths,
limitations and complexity are evaluated using literature where appropriate.
Important physical constraints and conserved quantities are identified,
and the approximations made are clearly stated. The purpose of the
calculation and desired results are clearly articulated.\tabularnewline
\hline 
Planning & The program to be written is broken into modules and functions that
can be designed, tested and debugged independently. A suitable and
efficient representation of the data, such as classes and data structures,
is chosen appropriately for the algorithm. Relevant libraries, software
packages and existing code are identified. The need for parallelization
is considered and incorporated into the design if necessary. An appropriate
language is chosen based on needs of the problem, ease of implementation,
maintainability and performance.\tabularnewline
\hline 
Implementation & The code can be easily understood and convinces the reader it works
through careful commenting, descriptive variable and function names
and validation of input. Coding standards are developed and obeyed
amongst the implementation team. Comments document the physics, are
in proportion to the complexity of the section, and identify input
and output to functions. Outside documentation thoroughly describes
how the program works and how to run it such that future users and
maintainers can easily work with the program. The code is efficient
but not at the expense of readability and maintainability. Professional
programming tools, e.g. version control and debuggers are used where
available.\tabularnewline
\hline 
Testing & Each element of the program is carefully tested separately and together
at each step. The program is verified on test cases with known solutions
identified in the planning process. Appropriate metrics are used to
analyze performance and identify need for optimization. Visualization
is used to provide insight into whether the algorithm is working.\tabularnewline
\hline 
Running & Initial conditions are chosen judiciously. Output is organized and
labeled and input parameters used in each run are recorded. Multiple
runs, if necessary, are automated efficiently through scripts.\tabularnewline
\hline 
Visualization & Visualization is used to gain intuition regarding the output and to
present final results in a compelling way. Important results are visualized
so that they require little effort to decipher and are appropriate
for the given data. Relevant tools are used to make the visualization.
Strengths and limitations of the choice of representation and alternatives
are discussed. Visualization, including physical objects, may also
be used to display program and data structures if necessary.\tabularnewline
\hline 
Numerical Analysis & The source and nature of all approximations made are identified and
their impact on the result discussed. The most significant sources
of error are carefully analyzed and estimates of the error are given;
ideally these are used to guide the algorithm, e.g. in refining the
discrete representation. Conditions for stability are identified and
quoted. The scaling of computational time with problem size is understood.
Reproducibility is verified for non-deterministic algorithms. Floating
point arithmetic is used carefully.\tabularnewline
\hline 
Physical Analysis & Results are compared to hypotheses. Adherence to physical constraints
(e.g. energy conservation) is verified. Possible improvements or alternate
implementations are identified. The results are cross checked with
alternative simulation strategies.\tabularnewline
\hline 
\end{longtable}
\par\end{centering}

\caption{\label{tab:Rubric}Rubric based on expert practice.}

\end{table*}

\section{Course design\label{sec:Course-design}}

\begin{table*}
\begin{longtable}{|l|>{\raggedright}p{2in}|>{\raggedright}p{1.5in}|>{\raggedright}p{1.5in}|>{\raggedright}p{1in}|}
\hline 
Week & Learning Objectives & Assignments & Content  & Additional skills\tabularnewline
\hline 
1-2 & Documenting code & Double pendulum & Discretization \textemdash{} ODEs \textemdash{} Integration & Group work\tabularnewline
 & Testing each component separately &  &  & \tabularnewline
 & Version control &  &  & \tabularnewline
\cline{1-4} 
3-4 & Dividing code into functions & 1D quantum mechanics & Eigensystems \textemdash{} Root finding \textemdash{} Shooting & Reading papers\tabularnewline
 & Identifying sources of error &  &  & \tabularnewline
\cline{1-4} 
5 & Selecting a model or algorithm & Model fitting & Optimization \textemdash{} Statistics & \multirow{2}{1in}{Consulting textbook/internet resources}\tabularnewline
 & Identifying improvements &  &  & \tabularnewline
\cline{1-4} 
6 & Identifying sources of approximation & Linear combination of atomic orbitals & Linear Algebra & \tabularnewline
 & Using visualization to analyze results &  &  & \tabularnewline
\cline{1-4} 
7-8 & Scripting & \multirow{2}{1.5in}{Ising model of magnetism} & \multirow{2}{1.5in}{Monte Carlo \textemdash{} Random Numbers} & \tabularnewline
 & Reproducibility &  &  & \tabularnewline
\cline{1-4} 
9-10 & Identify symmetries / conserved quantities & Time dependent Schrödinger equation & PDEs & \tabularnewline
 & Test physical acceptability of sol'n &  &  & \tabularnewline
\cline{1-4} 
11-12 & Identify existing packages & Laplace's equation  & \multirow{2}{1.5in}{Relaxation \textemdash{} Interpolation/Extrapolation} & \tabularnewline
 & Choose representation of data structures &  &  & \tabularnewline
 & Analyze stability/scaling &  & Finite Differences & \tabularnewline
 & Analyze and optimize performance &  & Finite Elements & \tabularnewline
\cline{1-4} 
ongoing & Develop a hypothesis & Final project &  & \tabularnewline
 & Choose language &  &  & \tabularnewline
 & Use visualization to present a compelling case &  &  & \tabularnewline
\hline 
\end{longtable}

\caption{\label{tab:CoursePlan}Course plan}
\end{table*}

Having created the rubric described in section \ref{sec:Rubric},
we set out to design a course in which students could learn the expert
practices identified in a structured manner. Following other studies\cite{rebbi2008},
we decided to adopt a project-based approach. Our overall vision for
the course was that students would complete a structured sequence
of projects to acquire the practices from the rubric in a scaffolded
manner. We proposed to spend the majority of the class time\textemdash $2\frac{1}{2}$
hours per week\textemdash with the students working in groups, guided
by the instructor and TA; we would introduce each class with a short
``micro lecture'' introducing each project and its associated concepts
and algorithms. We also anticipated that the students would create
a final term project where they would select and formulate the physics
problem to be solved themselves. 

First, we sought to distill the essence of the rubric further into
a small number of overall course goals for students to learn in the
course:\textemdash{}
\begin{enumerate}
\item Develop mathematical and computational models of physical systems.
\item Design, implement and validate a functioning code for such a model.
\item Understand the role of numerical analysis in formulating, designing
and interpreting computer models.
\item Use visualization and physical analysis to test hypotheses.
\item Connect theory, experiment and simulation.
\end{enumerate}
To identify possible projects, we surveyed 17 faculty from the Astronomy,
High Energy Physics, Cosmology and Condensed Matter Physics research
groups in the Tufts Physics and Astronomy department. For each of
the proposed projects\textemdash which we supplemented with some of
our own\textemdash we selected practices from the rubric that might
be particularly important. While good programming is a ubiquitous
requirement, for example, the Ising model is a suitable problem to
study random numbers and issues of repeatability. We then ordered
this collection of projects to determine those that could be completed
with little specialist knowledge to those requiring more, and identified
dependencies between projects. 

Following the backward course design approach\cite{wiggins2005understanding},
we selected a subset of projects that allowed us to span the full
list of practices and constructed the course plan shown in table \ref{tab:CoursePlan}.
Reading column-wise for each project is an estimate of the time required\textemdash 1
or two weeks\textemdash the practices we intended to focus on in the
project, the project title, the content of the project as viewed from
a traditional course, and any additional skills that may be required
for successful completion of the project. From this overall plan,
we developed project descriptions, available as supplementary material\cite{supplementary},
as well as brief ``micro-lectures'' to communicate both the content
(i.e. algorithms and concepts) and the Learning Objectives (i.e. the
professional practices). The problems themselves will be described
in more detail in the following section concerning our initial implementation.

\section{Implementation\label{sec:Implementation}}

\subsection{Student backgrounds\label{sub:Student-backgrounds}}

\subsubsection{Demographics}

An initial implementation of the course ran in the Spring semester
of 2015 and was ultimately completed by $n=22$ students out of 23
registrants. The course contained $7$ women and $14$ men. Some 13
were Physics or Astronomy majors, 8 were Computer Science majors with
some double majoring in both Physics and Computer Science, 3 were
graduate students in Physics and Education.

\subsubsection{Prior experience}

In this section, we discuss the prior experience of students in the
class. We first address the issue of programming language. The course
design presented above is intentionally language independent because
evidence from our interviews indicates that good programming practices
are largely independent of the language chosen. Moreover, language
selection for a given problem is an important skill. The language
chosen should be one used by professional physicists; within the general
family of languages familiar to the students; have good quality development
tools and have libraries or intrinsic functions to support useful
algorithms and visualization. 

\begin{table}
\begin{centering}
\begin{tabular}{cccccc}
\hline 
 & None & Intro. & Inter. & Adv. & Grad.\tabularnewline
\hline 
\hline 
Physics & 0 & 1 & 7 & 4 & 2\tabularnewline
CS & 4 & 6 & 3 & 1 & 0\tabularnewline
Math & 0 & 3 & 7 & 4 & 0\tabularnewline
\hline 
\end{tabular}
\par\end{centering}

\caption{\label{tab:Subject-preparation}Subject preparation.}
\end{table}

\begin{table}
\begin{centering}
\begin{tabular}{|c|c|ccc|}
\cline{3-5} 
\multicolumn{2}{c|}{} & \multicolumn{3}{c|}{Physics}\tabularnewline
\cline{3-5} 
\multicolumn{2}{c|}{} & Intro. & Inter. & Adv.\tabularnewline
\hline 
\multirow{4}{*}{CS} & None & 0 & 1 & 3\tabularnewline
 & Intro. & 1 & 3 & 2\tabularnewline
 & Inter & 0 & 2 & 1\tabularnewline
 & Adv.. & 0 & 1 & 0\tabularnewline
\hline 
\end{tabular}
\par\end{centering}

\caption{\label{tab:Subject-correlation}Correlations between subject preparation
in Physics and Computer Science.}
\end{table}

To aid this decision, and generally guide our preparation, we circulated
a brief survey before the course to assess student course preparation
in physics, computer science and math as well as programming language
familiarity and confidence in programming. The questions are listed
in Appendix \ref{sec:AppendixStudentBackground}. From the $14$ responses,
we coded student preparation in each subject as Introductory, Intermediate,
Advanced or Graduate depending on the highest level course taken by
the student. The courses General Physics 1 and 2, Calculus 2, Multivariable
Calculus and Introduction to Programming were classified as Introductory.
Modern Physics, Linear Algebra, Differential Equations and Data Structures
were classified as Intermediate. Any courses with higher numbers were
classified as Advanced. 

The results for subject preparation are shown in table \ref{tab:Subject-preparation}.
Most obviously, students were best prepared in Physics, but only slightly
less well prepared in Math. While most students had some familiarity
with Computer Science, it is very clear that preparation in this subject
was weaker than the other two subjects. A table of student numbers
sorted by both Physics and Computer Science preparation, displayed
in table \ref{tab:Subject-correlation}, shows that the majority of
students had an intermediate or better preparation in Physics and
only introductory or no preparation in computer science. A small minority,
however, had an intermediate or better preparation in in both subjects.
No students had taken advanced or graduate classes in both subjects.

\begin{table}
\begin{centering}
\begin{tabular}{ccc}
\hline 
Language & Familiar & Confident\tabularnewline
\hline 
\hline 
C++ & 10 & 8\tabularnewline
Mathematica & 9 & 4\tabularnewline
Python & 9 & 4\tabularnewline
Java & 6 & 3\tabularnewline
MATLAB & 5 & 3\tabularnewline
FORTRAN & 1 & 0\tabularnewline
\hline 
\end{tabular}
\par\end{centering}

\caption{\label{tab:Language-preparation}Language familiarity.}
\end{table}

Languages with which students expressed familiarity or confidence
are shown in table \ref{tab:Language-preparation}. The list is somewhat
abbreviated: exactly one student each expressed familiarity with Javascript,
Lua, Objective-C, Ruby, Scheme, SQL, Swift and Visual Basic. All respondents
self-assessed their own experience with programming as familiar or
better. Interestingly, \emph{all} students provided concrete examples
of programming tasks they had completed outside Introductory Computer
Science classes, referring to data analysis, simulations and programs
written for research purposes. 

With this information, we decided to use both \emph{Mathematica} and
Python for the course as these were both near the top of languages
students were already familiar with. Both of these languages are high
level, easy to learn and provide rich visualization tools; \emph{Mathematica}
also provides access to many useful algorithms. Because \emph{Mathematica}
is better suited to short programs and has the shallower learning
curve, it was used in the first half of the class; we then switched
to Python for the Ising model and subsequent projects, except for
the project on Laplace's equation. Additionally, we allowed students
to use other languages if they wished, though this option was only
utilized for the final projects. By giving students the experience
of using more than one language, we were able to explicitly compare
and contrast their strengths and weaknesses.

\subsection{Logistics}

\begin{table}
\begin{centering}
\begin{tabular}{cc}
\hline 
Preferred group size & No. respondents\tabularnewline
\hline 
\hline 
2 & 4\tabularnewline
3 & 6\tabularnewline
4 & 2\tabularnewline
5 & 0\tabularnewline
\hline 
\end{tabular}
\par\end{centering}

\caption{\label{tab:GroupSize}Preferred group size.}
\end{table}

In this section, we describe our approach to organizational matters.
The class met twice a week for $1\frac{1}{4}$ hour sessions. As noted
above, the typical class structure involved short presentations by
the instructor on the project physics background, a numerical analysis
topic, or one of the competencies identified in the rubric. \emph{Numerical
Recipes}\cite{Press} was recommended as a textbook, and was used
by students as a reference for standard algorithms. Support was provided
through portions of class time dedicated to student group work during
which the instructor and TA were available to help troubleshoot issues,
as well as traditional office hours. A \emph{Piazza} forum was also
provided for the course which received 2-3 posts per week. Questions
on the forum were typically answered quickly, on average within half
an hour. Students tended to use the forum to share links to helpful
resources, organize groups for the final project, and pose syntax
questions. 

A key issue in designing the class was the choice of group size. Following
generic advice, the first two projects listed below were executed
in teams of 4 or 5; project 3 was performed in pairs; project 4 was
performed in teams of 4. At this point in the semester, we circulated
a mid-semester survey with one question that asked students what they
thought the optimal group size was based on their experiences. We
had 12 responses, which are summarized in table \ref{tab:GroupSize}.
Students were also asked to explain the reasons for their selection,
which revealed two opposing effects: logistics, such as finding convenient
times to meet physically and synchronizing efforts favor smaller group
sizes while division of labor tends to favor larger group sizes. However,
students reported that in larger groups, there tended to be people
who didn't contribute significantly. We therefore concluded that groups
of 3 were preferable for this course and used this size for the remainder
of the semester with occasional groups of 4 where required to accommodate
the whole class.

Another important matter is the issue of grading in a team-based class.
As described above in section \ref{sec:Course-design}, the instructors
gradually scaffolded in the competencies during the course and this
was reflected in the grading. For example, the first project was graded
primarily on the quality of the code as well as successful completion
of the project. A key mechanism we established to provide accountability
was to require short individual self-assessments to be completed after
each project online through Tufts' Course Management System (Sakai).
The questions asked are listed in appendix \ref{sec:SelfAssessment}.
These submissions allowed the instructors to understand who had done
what as well as where students felt their own work stood in comparison
to the grading criteria. Project submissions were graded as a team,
but each student received an individual grade and feedback form\textemdash collated
and emailed automatically by a \emph{Mathematica} script\textemdash including
their own self-assessment as a component. The instructor occasionally
modified grades if there were sufficient evidence that an individual
had failed to contribute meaningfully to a project, though this was
used very sparingly. 

An overview of the projects, as well as the project descriptions themselves,
is provided in supplementary material\cite{supplementary}. A final
project, performed in parallel with the last third of the class, was
intended as the summative assessment, allowing students to attempt
a problem of their own choosing and replacing the final exam of a
typical class. A list of project ideas was provided, largely connecting
with TJAs research in Condensed Matter Physics, but included the option
to formulate a problem that connected with student's research field.
Some students chose to work individually, others in teams of up to
four. The class concluded with a mini research symposium where each
group gave a short five minute presentation showcasing their findings.
Each student was required to fill out a self-assessment different
from the other projects, asking instead how the project demonstrated
their professional practice in each competency.

\subsection{Student work}

\begin{figure}
\includegraphics[width=1\columnwidth]{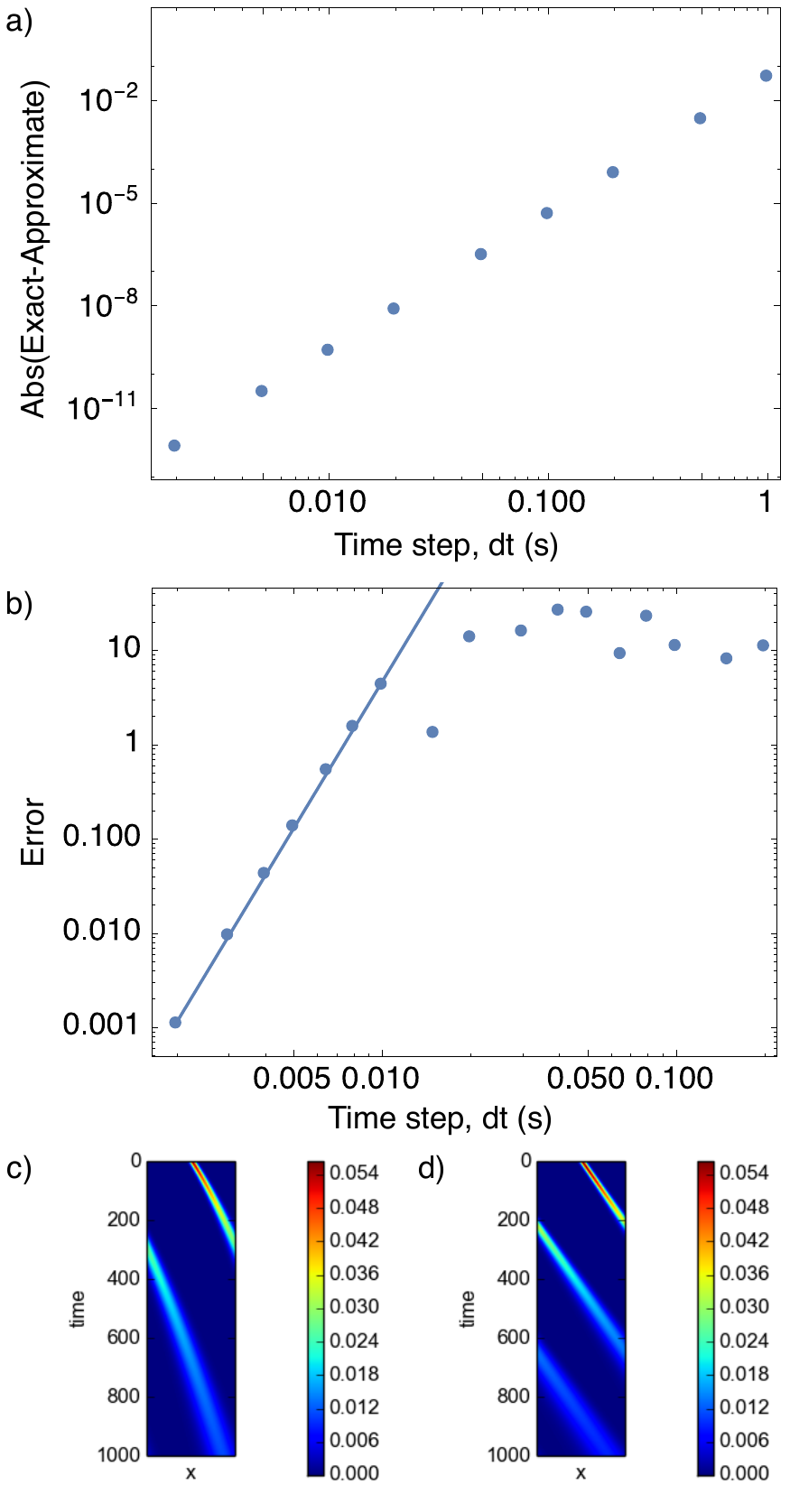}

\caption{\label{fig:StudentWork}Examples of student-generated visualizations.
(a) Finding the dependence of the error in a Runge-Kutta integration
of the simple pendulum; (b) a similar plot for the double pendulum.
(c) and (d) Comparing the performance of two different algorithms
for integrating the TISE. }
\end{figure}

We present two brief case studies of student work that illustrate
some of the ways groups creatively engaged with the projects and used
visualization to analyze the performance of their program. The first
example comes from project 1, the Double Pendulum. In this project,
each group implemented a different algorithm, and the project was
performed in two stages: first, students used their assigned algorithm
to solve the simple harmonic oscillator equations and presented their
findings to the class, afterwards they extended their code to solve
the double pendulum. The numerical analysis concepts of Error, Order
and Stability were introduced in class, but only the performance of
the Euler algorithm was discussed in detail. The group that used the
Runge-Kutta algorithm spontaneously conducted an analysis of the discrepancy
between their numerical solution and the analytical solution as a
function of time-step. They presented the graph shown in fig. \ref{fig:StudentWork}(a)
to the class after the first part of the project, and used the \texttt{FindFit}
function of \emph{Mathematica} to verify that the error was $\propto\left(\Delta t\right)^{4}$
as expected. In the second part of the project, they created a similar
plot {[}fig. \ref{fig:StudentWork}(b){]}, comparing the solution
at a specified $\Delta t$ to that with much smaller $\Delta t=0.0005$.
The graph was annotated in their notebook by the comment,
\begin{quotation}
\emph{``show max error and the convergence fit for small dt, which
is actually of order 5 in this case. Note that the error is considerably
larger, even at small dts, than the more well-behaved single pendulum,
and also that the chaos in the system is very evident at larger dt''}
\end{quotation}
indicating that this group had appreciated that the Runge-Kutta algorithm
was performing as expected in some regime of $\Delta t$, but failing
in others, and that the highly nonlinear double pendulum equations
were more challenging to solve. 

The second example comes from project 6, the Time Dependent Schrödinger
Equation, in which students solved a number of classic 1D problems.
For their report, most groups presented snapshots of the wavefunctions
at different times. One group, however, utilized a kymograph representation
to show both spatial and temporal evolution. Two examples taken from
their report are shown in fig. \ref{fig:StudentWork}(c) and (d),
both showing the propagation of a gaussian wavepacket as a function
of time. Periodic boundary conditions are enforced, so the wavepacket
exits from the right and re-enters from the left during the simulation.
The two figures represent different algorithms, fig. \ref{fig:StudentWork}(c)
is a non-unitary cell-centered finite difference scheme; fig. \ref{fig:StudentWork}(d)
is a unitary scheme. Paired with these figures was the text,
\begin{quotation}
\emph{``By applying periodic boundary conditions, we see the particle
exit the frame on the right and reappear on the left, as expected.
The results from the two different algorithms are shown in Figure
X. Notice how the first algorithm gives undesirable results, such
as dissipation and a non-constant velocity of the wavepacket. The
second algorithm, however, gives the expected results.''}
\end{quotation}
showing that the students selected this representation to make such
comparisons straightforward. The wavepacket in fig. \ref{fig:StudentWork}(c)
clearly changes velocity as the slope of the line changes. The students
went on to use this representation to verify time-invariant behavior,
i.e. eigenfunctions, as well as reflection from a barrier and crystal.

\section{Analysis\label{sec:Analysis}}

In this section, we present feedback on the initial implementation
of the course from several sources. An informal feedback session was
held during class and students completed the usual course evaluation
required by the university. Furthermore, we asked the experts interviewed
to develop the rubric in section \ref{sec:Rubric} to regrade student
work to determine whether it aligned with their conception of professional
practice.

\subsection{Informal feedback}

Students in the final class were asked to collaboratively construct
ideas for things to do differently in future classes. First among
these were the course requirements, which were kept deliberately minimal
in the first iteration. Students thought that the requirements should
be Modern Physics due to the presence of Quantum Mechanics projects,
an intermediate math class such as Linear Algebra and the second Computer
Science class, i.e. Data Structures. We note that many, but not all,
students would have met these requirements given the backgrounds presented
in section \ref{sub:Student-backgrounds}. A second important issue
was the level of course credit and class time. The course was initially
offered as a Special Topics class worth 1 course credit (equivalent
to 3 Semester Credit Hours); students felt that $1\frac{1}{3}$ or
$1\frac{1}{2}$ might be more appropriate given the workload, similar
to a lab course, and that additional class time e.g. through an associated
recitation would be valuable. When prompted by the instructor, students
preferred to increase the course credit rather than delete projects.
It was felt that an opportunity to cross-examine other groups work
would be helpful, first by having students upload their work to a
repository, and second by incorporating an after project de-brief.
Finally, students requested more structure and scaffolding in the
initial projects, including internal deadlines and clearer stages.
Following the class, several students noted that they had continued
to work on projects initiated in the class; we are aware of at least
one project that resulted in a journal publication.

\subsection{Course evaluations}

The course evaluation was completed by $15$ students. Tufts course
evaluation questions use a 5 point scale (Very poor\textemdash Less
than satisfactory\textemdash Satisfactory\textemdash Very good\textemdash Excellent).
Students' overall evaluation of the course was a mean of $4.47$ corresponding
to Excellent and compared to a departmental average for undergraduate
major courses of $4.1$, weighted by enrollment. The mean response
to the question ``How would you rate the success of the course in
accomplishing its objectives as stated on the course syllabus'' was
$4.40$ and ``How would you rate the use of out-of-class activities
to promote your learning'' yielded a mean of $4.20$, indicating
that the project-based approach was well received. Of mild concern
is that students' responses to a question on the use of in-class time
received a lower mean score of $3.73$, suggesting that some reorganization
of this aspect might be necessary. 

Written response questions provided more detail to these views. In
response to the question, ``In what ways has this course made you
think differently or more deeply?'', most students commented on the
subject matter, for example,
\begin{quote}
\emph{``this class gave me a glimpse of many different types of interesting
physics problems. The range of problems tackled during the class gave
me a sense of the range of applications of computation in physics''}

\emph{``The projects were felt 'real'. They were non-trivial and
felt like the kinds of problems a real computational physicist might
work on''}
\end{quote}
or the problem-solving aspects,
\begin{quote}
\emph{``It was really amazing to get to learn about this approach
to solving problems\textemdash how to approximate solutions, how to
tell how accurate the approximation is, and how to make physics so
much more accessible with computers''}

\emph{``The class made me think more deeply about computation. Instead
of being content to write code that gets something done, I think about
how the code is getting something done and what methods will be best
to reduce error, runtime, etc. This is an extremely useful and marketable
class. I could have used this ages ago.''}
\end{quote}
but also group work,
\begin{quote}
\emph{``Combining coding skills with communication skills was a new
and different task for me.''}

\emph{``I am better at working in groups and I now have an idea of
what it is like to do computational physics as a profession.''}
\end{quote}
In response to ``What aspects of this course worked best to facilitate
your learning?'', students universally mentioned the projects, which
were \emph{``super interesting and well thought out''}. However,
positive experiences with group work was also a theme, in spite of
the challenges, 
\begin{quote}
\emph{``Group work was great. I say this in spite of my complaints
about certain group members. Students were forced, in these projects,
to go out and discover how to make a certain approach work. This led
to strong engagement and lasting impression.''}

\emph{``Different groups worked together in different ways, and not
all of them were great. But I feel like I really know and like a ton
of the people in the class because we were together through the hardest
of times. I also feel like I was able to learn a lot of different
things from everyone I worked with.''}
\end{quote}
The final question on how to improve the course provoked a great deal
of responses. The vast majority of these repeated topics in the in-class
discussion, including the need for prerequisite courses, additional
class time and managing group size. Many of the remainder touched
on more delicate themes about group work. Challenges identified included
that \emph{``there was a TON of variance in ability'' }as well as
others referring to differing levels of effort put in by some students.
One student felt that,
\begin{quote}
\emph{``{[}The instructor should{]} more clearly adjust grades of
students who don't participate equally in groups. I have a sense that
it was 2-3 students who were consistently bad group members, but it
was a definite problem. Without the grade incentive, I honestly think
that there's a portion of students who will just not do the work and
coast along.''}
\end{quote}
\noindent Another student suggested additional structure for group
work,
\begin{quote}
\emph{``{[}People{]} need to be taught time management if they are
going to be working in groups. In order to do this, I think that you
should have intermediate deadlines for the first few projects (not
just the first one), and you should require that students turn in
a work plan (including internal deadlines) before they do anything
else for each project. Without an authority endorsing this sort of
behavior, responsible group members are unlikely to be able to hold
other group members accountable.''}
\end{quote}
\noindent Several students also mentioned the need for additional
supporting resources,
\begin{quote}
\emph{``I would make sure those of us with less experience have clear
resources for information. It seemed like I had no way of figuring
out what to do or where to turn sometimes.''}

\emph{``{[}I'd like to see{]} Reading recommendations to get necessary
mathematical background''}

\emph{``I would provide some more background information for people
who may not have extensive math or programming background. The }Mathematica\emph{
and python tutorials that were posted were really helpful, so more
online resources or 'crash course' type things for some of the computational
and mathematical methods we used would have been great to get everyone
on the same page.''}
\end{quote}
Overall, feedback from students in both formal and informal contexts
suggests that the course was a success, that this form of pedagogy
is effective, if intensive, but needs additional support to accommodate
students with different course backgrounds as well as more structure
in managing group dynamics.

\subsection{Post course grading study}

From the initial implementation of the course, we wished to determine
whether student work produced in the class satisfied the views on
professional practice held by the experts whom we interviewed to create
the rubric. Moreover, we wished to establish whether there was evidence
that these were acquired during the class itself. 

\begin{table}
\begin{centering}
\begin{tabular}{|c|ccccc|ccccc|}
\hline 
 & \multicolumn{5}{c|}{Submission A} & \multicolumn{5}{c|}{Submission B}\tabularnewline
\cline{2-11} 
 & 1 & 2 & 3 & 4 & 5 & 1 & 2 & 3 & 4 & 5\tabularnewline
\hline 
\hline 
Phys. Trans. & F & F & F & P & P & G & G & G & G & G\tabularnewline
Planning & \textemdash{} & G & G & G & F & G & G & G & G & G\tabularnewline
Implementation & \textemdash{} & F & F & G & F & G & F & G & G & G\tabularnewline
Testing & F & \textemdash{} & \textemdash{} & G & \textemdash{} & G & \textemdash{} & \textemdash{} & G & G\tabularnewline
Running & F & P & \textemdash{} & P & F & G & G & G & G & G\tabularnewline
Visualization & P & F & F & \textemdash{} & \textemdash{} & G & G & G & G & G\tabularnewline
Num. Ana. & P & \textemdash{} & P & P & P & F & G & G & G & G\tabularnewline
Phy. Ana. & P & P & P & P & \textemdash{} & G & G & G & G & G\tabularnewline
\hline 
\end{tabular}
\par\end{centering}

\caption{\label{tab:Comparison-of-expert}Comparison of expert grading of two
projects. }
\end{table}

To test inter-grader consistency, we selected two submissions from
project 5 that the we felt displayed a large difference in quality,
receiving a C grade and an A grade respectively. We asked the experts
to grade these two projects, for each competency assigning a label
from the categorizations Good\textemdash Fair\textemdash Poor as well
as Insufficient Evidence. The results are shown in table \ref{tab:Comparison-of-expert}.
Submissions A and B were clearly distinguished from one another in
quality by the graders, but the degree of consistency is different
between the two. To find out why, we held a panel with four of the
experts in attendance and invited them to discuss their rationale
where there seemed to be discrepancies. In discussion, it seemed that
most of the difference between graders could be attributed to whether
or not they looked at the code to make their distinction or just looked
at the report. For example, grader 4 felt that the quality of the
implementation for submission A was good, having looked at the code
and noted extensive commenting. The other graders had given more weight
to the quality of the report. The experts all agreed that submission
B represented work of a high quality. In the panel, we asked all the
graders whether they felt that submission B accorded with their own
vision of expert practice and they answered affirmatively. We therefore
established that at least some students displayed evidence of expert
practice as articulated by our rubric, and determined additional guidance
necessary to promote inter-grader consistency. 

\begin{figure}
\includegraphics[width=3.3in]{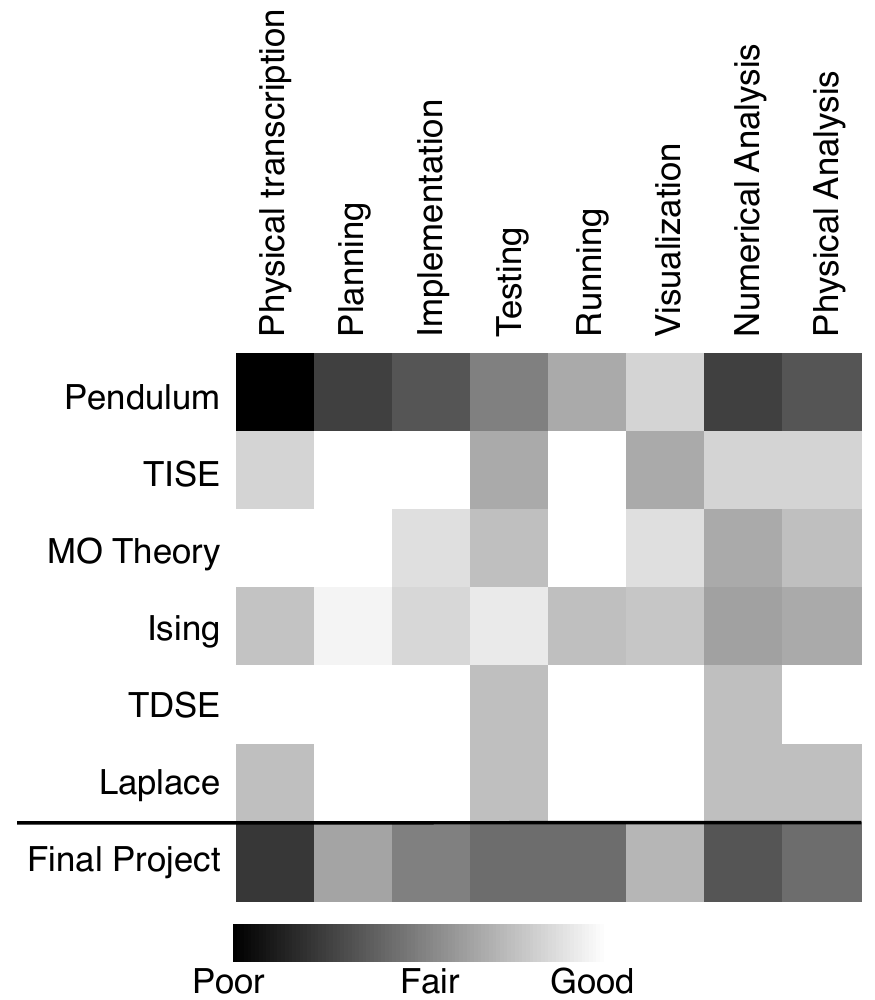}

\caption{\label{fig:Post-course-grading-study}Post-course grading study. Project
submissions were regraded after the class by the expert participants}

\end{figure}

To track the work longitudinally, we amended our grading instructions
to require examination of the code and asked the graders to grade
other projects from the course. Project 3 involving data fitting was
not graded because it was substantially shorter and more prescriptive
than the other projects. The projects were stripped of information
that identified their ordering, and randomly assigned to different
graders; each submission was graded by a single grader, and each competency
was graded on the same Good\textemdash Fair\textemdash Poor scale
above. In total, 24 submissions from possible total of 49 were regraded. 

From these, a mean grade was calculated for each project that is displayed
in the visualization in fig. \ref{fig:Post-course-grading-study}.
This is intended to provide a snapshot of how well students acquired
the competencies during the class. Some trends are immediately clear:
happily, it appears that after the initial project, the quality of
work rapidly improved, and continued to improve up to the last class
project (Laplace's equation). The very swift improvement after the
first project suggests that the students already possessed relevant
abilities\textemdash whatever they were\textemdash to do this kind
of work, but needed to find out how to apply them. It is possible
that the presence of the rubric helped this. 

There are differences between the skills: some such as Planning, Implementation
and Running seem to have been picked up without trouble. Visualization
improved substantially later in the class. Two skills, however, Testing
and Numerical Analysis seem to have presented more difficulty. It
is important to be careful in interpreting this: it is not necessarily
the case that students failed to test their code adequately, though
this may be true, but rather that the evidence of testing was insufficiently
detailed to the graders. Physical transcription and Physical analysis
were also somewhat inconsistent, suggesting that the more abstract
competencies are more challenging for students to acquire. In future
iterations of the class, additional scaffolding and emphasis should
be introduced to address these competencies. It is also clear that
the final projects were substantially weaker, in the eyes of the graders,
than most of the class projects. In part, this is because the write-ups
for these projects were kept very short; it is also inevitable that
formulating and executing a project is a much more complex task than
solving a prescribed problem. While the instructors were very pleased
with the results of the final projects, it appears that in future
classes additional time, support and reporting requirements should
be introduced to improve the final projects.

\section{Conclusion\label{sec:Conclusion}}

We present a project-based Computational Physics Course that was designed
around expert practice obtained from interviewing faculty and distilled
into a rubric that summarizes identifiable features of this practice.
A successful initial implementation of the course, together with extensive
post course analysis, was also presented as well as matters to consider
in future iterations of the course. By having the expert interviewees
grade student work produced in the class, it is clear they found evidence
of excellent practices that were acquired by students in the class.
It is furthermore clear that the rubric can be used as a tool to separate
a strong submission from a weak one. 

Our work builds upon previous studies of Computing in Physics Undergraduate
Education and proposed course structures by providing an concise version
of what constitutes expert practice in this discipline, a tool to
measure it and a reference implementation. Due in part to the small
size of this study, there are remain numerous avenues for further
investigation. For instance, a key limitation is that expert practices
were self reported and not directly observed. Even so, our interviews
complicate simplistic models of expert problem solving originally
formulated for introductory classes. At the very least, such models
may simply be inapplicable to complex research-oriented problems such
as those attempted by students in this class. There is a clear need
for more realistic models, of which that presented in fig. \ref{fig:FinalModel}
is likely to prove a crude initial sketch. Of course, a scaled-up
multi-class multi-university study on project-based approaches to
Computational Physics is likely to prove deeply informative, particularly
since similar techniques are now being tried in introductory classes.
Given the central importance of group dynamics to this strategy, which
we have shown to be both challenging and rewarding, micro-level approaches
such as video observation and interviews with students might help
untease how students participate in group work in this context. Our
work also provides a valuable example of backwards course design that
may be applicable in physics to domains other than computation, such
as laboratories or graduate classes focussed on preparing students
for research.

While our study has shown that students taking a single course on
Computational Physics \emph{can} acquire expert practice, the broader
question of the role of Computation in Physics Education remains open.
Is it best to have a small number of courses such as the one presented
that focus primarily on Computation, is it better to intersperse Computation
into many classes, or is a mixed approach needed? Identifying strategies
that produce graduates and PhD students well prepared to perform computational
work that meets the needs of future research, industry and other careers
is surely of considerable benefit to the community of Physicists. 
\begin{acknowledgments}
\emph{The authors thank Prof. Michelle Wilkerson as well as Sheryl
Barnes and Dona Qualters for much invaluable expertise in helping
us design and conduct the study, and Profs. Roger Tobin and David
Hammer as well as Anna Phillips for providing thoughtful advice on
the manuscript. We are grateful to Tufts University for a Tufts Innovates!
seed grant to create and implement the course. CJB acknowledges funding
from the Graduate Institute For Teaching (GIFT) program at Tufts University
that partially supported his participation. TJA is supported by a
Cottrell Scholar Award from the Research Corporation for Science Advancement. }

\appendix
\end{acknowledgments}

\section{Appendix: Expert interview questions\label{sec:Appendix-A}}

The following sequence of questions was asked in the expert interviews. 
\begin{enumerate}
\item Can you think of a problem in your research area that might be suitable
for an undergraduate computational physics course?
\item Could you explain the physics of the problem?
\item What are the possible computational approaches to solve this problem?
\item What are the advantages and disadvantages of these approaches?
\item Can you tell me what you see as the issues involved in transforming
a physics problem into an algorithm or set of equations to solve?
\item Thinking about the problem from earlier, how would you go about planning
a computer program to solve this problem?
\item Could you sketch a flowchart for this program?
\item I'm going to show you a page of code. Tell me what you like about
the programming, and what things would you change?
\item What do you think constitutes good programming?
\item What sort of visualization strategies do you think are useful? Which
do you use and what challenges do you come across?
\item What sort of visualization might you use for the example problem?
\item Using numerical analysis strategies, for example thinking about error,
order, and stability, how would you assess the numerical performance
of your algorithm?
\item More generally, what types of numerical analysis do you use to analyze
the algorithms you employ in your work?
\item How would you use the results of your hypothetical program to perform
a physical analysis of the system you are modeling?
\item What else might you consider while trying to come to physical conclusions
from a set of computational results?
\end{enumerate}

\section{Appendix: Student background survey questions\label{sec:AppendixStudentBackground}}

This short survey was circulated before the course. 
\begin{enumerate}
\item What Physics Courses have you taken?
\item What Computer Science or Applied/Numerical Math courses have you taken?
\item What Computer Languages have you used? Which do you feel confident
in?
\item Describe your overall experience and familiarity with programming.
Feel free to cite projects you've undertaken.
\end{enumerate}

\section{Appendix: Individual self assessments\label{sec:SelfAssessment}}

Self assessments were required after each project. 
\begin{enumerate}
\item Describe your contribution to the project. Identify things that you
yourself did. 
\item Overall, what grade would you give to your own contribution to the
project? \emph{(A\textemdash Mastery. I think I did this to a professional
level; B\textemdash Solid understanding. I got this, though there
may be still residual mistakes; C\textemdash Progress. I'm still working
on learning this.)}
\item How well did your team achieve the goals of the project? Explain briefly
each member's contribution. Identify any challenges your team faced
and how you overcame them. 
\item Overall, what grade to your team's project submission as a whole?
\emph{(A\textemdash Mastery. I think I did this to a professional
level; B\textemdash Solid understanding. I got this, though there
may be still residual mistakes; C\textemdash Progress. I'm still working
on learning this.)}
\item Did your team do anything over and above that required in the project
description?
\item If you have other comments on your group's project, please write them
here.
\end{enumerate}

\end{document}